\documentstyle[12pt,epsfig,rotating]{article}
\begin{document}
\baselineskip=18pt


\centerline{\bf The pion and kaon electromagnetic form factors}
 
\vspace{0.15in}
 
\centerline{J. Lowe$^a$ and M. D. Scadron$^b$}

\noindent $^a$Physics Department, University of New Mexico, 
Albuquerque, NM 87131
 
\noindent $^b$Physics Department, University of Arizona,
Tucson, AZ 85721
 
\vspace{0.2in}
 
\begin{abstract}
\noindent We use recent data on $K^+\rightarrow\pi^+e^+e^-$, together
with known values for the pion form factor, to derive the
kaon electromagnetic form factor for $0<q^2<0.125~{\rm (GeV/c)}^2$.
The results are then compared with predictions of the Linear $\sigma$ 
Model, a quark-triangle model and Vector Meson Dominance. The first
two models describe the data at least qualitatively, but the simple
Vector Meson Dominance picture gives a detailed quantitative fit to
the experimental results.
 
\vspace{0.15in}
 
\end{abstract}
  
\vspace{0.5in} 

\noindent{\bf 1. Introduction}
 
\vspace{0.25cm}
 
The pion electromagnetic form factor, $F_{\pi}(q^2)$, is well studied
experimentally. Many measurements, for both positive and negative 
$q^2$, have been reported in the literature \cite{radii}. Additional
information comes from the pion charge radius, which is related to
the slope of the form factor at $q^2=0$ (see \cite{PDG} for a summary
of measurements of the pion charge radius). By contrast, much less
is known for the kaon form factor, $F_K(q^2)$. There are some
measurements \cite{radii} for negative $q^2$, but no data exist for
$q^2>0$.
 
Information on $F_K$ can be deduced indirectly from experimental
data on the decay $K^+\rightarrow\pi^+e^+e^-$, such as that 
provided by the recent high-statistics Brookhaven experiment 
E865 \cite{865}. The amplitude for this decay was measured for $q^2$
up to 0.125 (GeV/c)$^2$, the maximum allowed by the kinematics of this
kaon decay. However, this amplitude does not give $F_K$ directly, but
rather the difference $F_K(q^2)-F_{\pi}(q^2)$. Since $F_{\pi}(q^2)$ is 
relatively well known, then, this decay is a source of information on 
$F_K$.
 
The Linear Sigma Model (L$\sigma$M) of Gell-Mann and Levy 
\cite{gml,MDS_FF,mike2} has proved to be fruitful in describing 
low-energy properties of mesons. In a recent theoretical paper 
\cite{MDS_FF}, the L$\sigma$M, and some related models, were used
to study meson electromagnetic form factors. The authors restricted
their comparison of $F_{\pi}$ and $F_K$ with experiment to a check
of the slopes at $q^2=0$, using the measured pion and kaon charge
radii. Here, we extend the comparison of these models with
experiment, and we examine their success in reproducing $F_K(q^2)$
for $0<q^2<0.125~{\rm (GeV/c)}^2$.

In section 2, we extract the kaon form factor from the experimental
data for $K^+\rightarrow\pi^+e^+e^-$. In section 3, we derive both
the pion and kaon form factors from the L$\sigma$M. In section 4, we
introduce the quark-triangle (QT) model and compare the QT
predictions with experiment and in section 5, we give the 
Vector-Meson-Dominance (VMD) model predictions. In section 6, we 
examine the predictions of these models for the slopes of the
form factors at $q^2=0$. Finally, section 7 summarises our results.
 
\vspace{0.5cm}
 
\noindent {\bf 2. The decay $K^+\rightarrow\pi^+e^+e^-$ and the kaon
form factor}
 
\vspace{0.25cm}
 
The decay $K^+\rightarrow\pi^+e^+e^-$ has been studied theoretically
for many years. Already in 1985 it became clear \cite{es} that
the process is dominated by the ``long-distance" (LD) terms, in
which a virtual photon is radiated by either the pion or the kaon.
However, it was not until the detailed data of experiment E865
\cite{865} became available that a convincing description of
both the scale and the $q^2$ dependence of the amplitude was found
\cite{bels}.
 
\begin{figure}[t]
\hspace{0.5in}
\epsfig{file=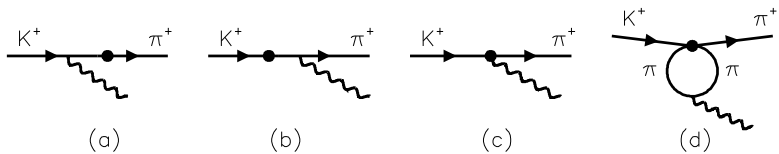,width=7.5cm}
\vspace{-1.6in}
\caption{\label{fig.1} Graphs for $K^+\rightarrow\pi^+e^+e^-$. 
(a) and (b) are
long-distance graphs, (c) is a short-distance graph and (d) is the
pion loop term. In each graph, the blob denotes the weak 
(strangeness-changing) vertex. The (off-shell) photon converts
to $e^+e^-$.}
\end{figure}

Burkhardt {\it et al.} \cite{bels} considered four contributions
to the amplitude, depicted in figure \ref{fig.1}. The LD terms are
those in figure \ref{fig.1}(a) and (b). These two graphs are related
to the pion and kaon form factors as shown below. Figure 
\ref{fig.1}(c) represents all short-distance
(SD) terms. These were already known in \cite{es} to be
small, and subsequent work \cite{ddg} has shown them to be smaller
than previously believed. Therefore here, as in ref. \cite{bels},
we neglect the SD contribution from figure \ref{fig.1}(c). Figure 
\ref{fig.1}(d) is a ``pion loop" term, first discussed by Ecker
{\it et al.} \cite{piloop}. Its contribution is small, but it gives
a characteristic shape to the $q^2$ dependence of the amplitude.
As in \cite{bels}, we take this term directly from
\cite{piloop}. Burkhardt {\it et al.} \cite{bels} gives a more
detailed discussion of the contributions to the amplitude.
 
The pion and kaon form factors enter {\it via} the graphs of
figure \ref{fig.1}(a) and (b), which give the LD amplitude 
\cite{bels} 
\begin{eqnarray}
\mid A_{LD}(q^2)\mid=e^2\bigg|
\frac{\langle\pi^+\mid H_W\mid 
K^+\rangle}{m^2_{K^+}-m^2_{\pi^+}}\bigg|~~\bigg|
\frac{F_{\pi^+}(q^2)-F_{K^+}(q^2)}{q^2}\bigg|.
\label{eq.1}
\end{eqnarray}
 
\noindent In the numerical calculations below, we take the value
of the weak matrix element $\langle\pi^+\mid H_W\mid K^+\rangle$
from \cite{ls}:
 
$$\mid\langle\pi^+\mid H_W\mid K^+\rangle\mid =(3.59\pm 0.05)
\times 10^{-8}~{\rm GeV}^2.$$

\noindent Adding the pion loop amplitude, figure \ref{fig.1}(d),
from Ecker {\it et al.} \cite{piloop}, we obtain

\begin{eqnarray}
A(q^2)&=&A_{LD}(q^2) + A_{\pi loop}(q^2)
\nonumber \\
 &=&e^2\bigg|\frac{\langle\pi^+\mid H_W\mid K^+\rangle}
{m^2_{K^+}-m^2_{\pi^+}}\bigg|~~\bigg|
\frac{F_{\pi^+}(q^2)-F_{K^+}(q^2)}{q^2}\bigg|+
A_{\pi loop}(q^2)
\label{eq.2}
\end{eqnarray}
 
\noindent from which

\begin{eqnarray}
\mid F_K-F_{\pi}\mid = \frac{q^2(m^2_{K^+}-m^2_{\pi^+})}
{e^2\mid\langle\pi^+\mid H_W\mid K^+\rangle\mid}
\bigg[A(q^2)-A_{\pi loop}(q^2) \bigg].
\label{eq.3}
\end{eqnarray}

To apply equation (\ref{eq.3}), we need experimental values 
for $A(q^2)$
and $F_{\pi}(q^2)$. For the former, we use data from Brookhaven E865
\cite{865}. Their amplitude $f(q^2)$ is related to our $A(q^2)$ by
 
\begin{eqnarray}
\mid A(q^2)\mid=f(q^2)\frac{G_F\alpha}{4\pi}
\label{eq.4}
\end{eqnarray}
 
\noindent where $G_F$ is the Fermi constant and $\alpha$ is the fine
structure constant. 

\begin{figure}[h]
\hspace{1.0in}
\epsfig{file=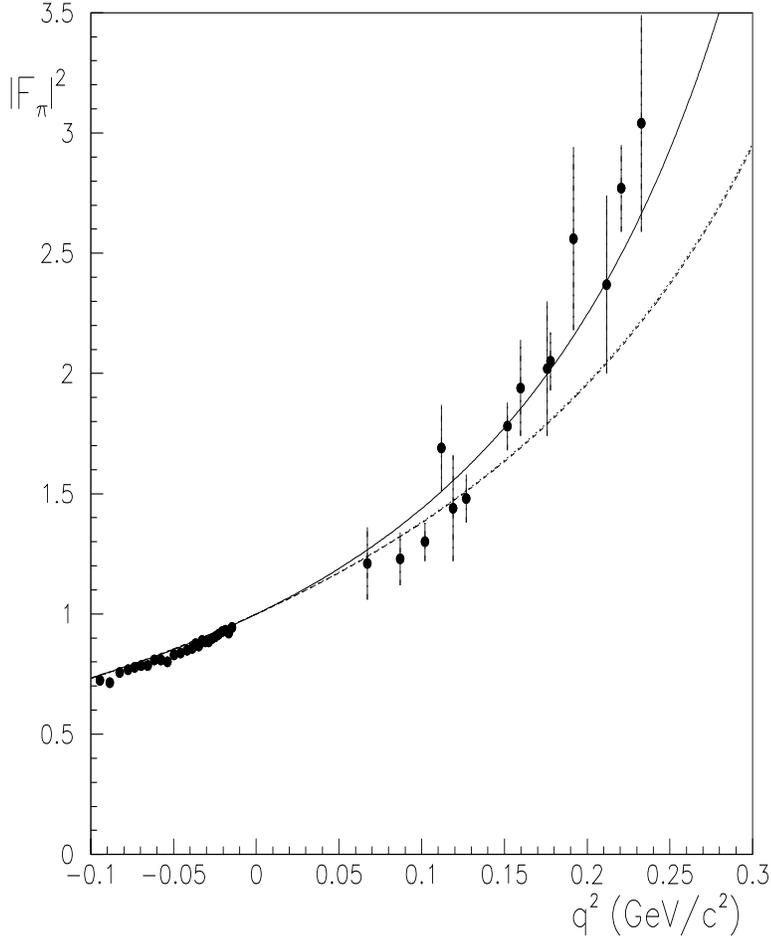,height=14.0cm,width=11.0cm}
\caption{\label{fig.2} Pion electromagnetic form factors squared.
The points are the experimental data. The curves are for VMD
(solid line), L$\sigma$M (dashed line) and QT (dotted line). The
dashed and dotted lines are practically indistinguishable in
the figure.}
\end{figure}

The experimental values of $F_{\pi}$ from \cite{radii}
are not in general at precisely the required values of $q^2$.
However, the data, which are plotted in figure \ref{fig.2}, are
well described by the VMD model using a rho-meson pole:

\begin{eqnarray}
F_{\pi}(q^2)_{VMD}= \frac{m^2_{\rho}}{m^2_{\rho}-q^2}
\label{eq.5}
\end{eqnarray}

\noindent for the region of $q^2$ of interest. This is shown by the
solid line in figure \ref{fig.2}. 
Therefore we use equation (\ref{eq.5}) to calculate the 
required values of $F_{\pi}$. We emphasise that the values we
calculate this way are essentially experimental; the VMD prediction
is used basically as an interpolating function between the
measured experimental points.

The relative sign of $A_{LD}$ and $A_{\pi loop}$ was already
established in \cite{865} and \cite{bels}. To derive
$F_K$ from equation (\ref{eq.3}), we need also to determine the
sign of
$F_K-F_{\pi}$. To do so, we observe that in all models discussed
below, as well as 
in the available data for $q^2<0$, $F_K$ differs less from unity
than does $F_{\pi}$, i.e.
$$\mid F_K-1\mid ~<~\mid F_{\pi}-1\mid.$$
This defines the required sign, giving
 
\begin{eqnarray}
F_K=F_{\pi}-\frac{q^2(m^2_{K^+}-m^2_{\pi^+})}
{e^2\mid\langle\pi^+\mid H_W\mid K^+\rangle\mid}
\bigg[A(q^2)-A_{\pi loop}(q^2)\bigg].
\label{eq.6}
\end{eqnarray}

\begin{figure}[h]
\hspace{1.0in}
\epsfig{file=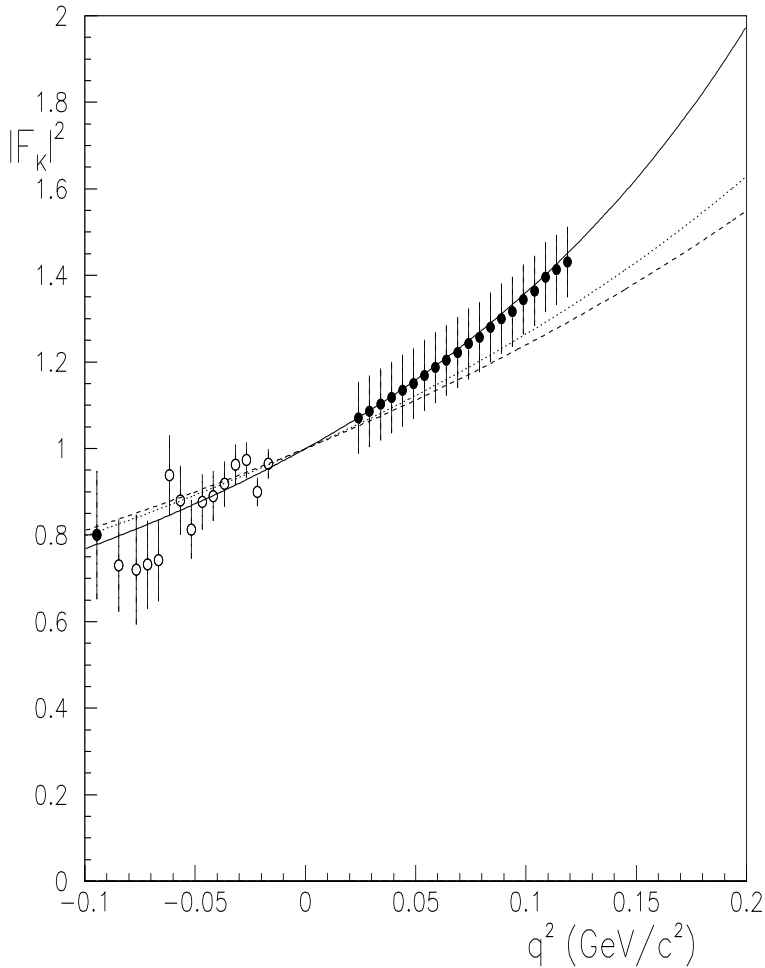,height=14.4cm,width=11.0cm}
\caption{\label{fig.3} Kaon electromagnetic form factors squared.
The solid points
are the experimental data from the present analysis and the circles
show the previously existing data. The curves are for VMD 
(solid line), L$\sigma$M (dashed line) and QT (dotted line).}
\end{figure}

The extracted values of $\mid F_K\mid^2$ are listed in table 
\ref{tab.1} and plotted in figure \ref{fig.3} which also shows
previous data for $q^2<0$. The errors in our values of $F_K$
arise from experimental errors in the amplitude $A(q^2)$ for 
$K^+\rightarrow\pi^+e^+e^-$ and the error in the weak matrix
element $\langle\pi^+\mid H_W\mid K^+\rangle$, but predominantly
from the errors on the experimental values of $F_{\pi}$. The VMD
model, equation (\ref{eq.5}), gives a reasonably unambiguous value
for
$F_{\pi}$. However, its use is only justified to the extent that
it agrees with the experimental points in figure \ref{fig.2}. The
weighted RMS deviation of the positive-$q^2$ experimental points
in figure \ref{fig.2} from the VMD prediction is 0.0837, and we
take this as the error in $\mid F_{\pi}\mid^2$.

\begin{center}
\begin{table}
\caption{\label{tab.1} Experimental values for the kaon form
factor from the present analysis.}
\vspace{0.3cm}
\center{
\begin{tabular}{|c|c||c|c|}   
\hline
$q^2$ & $\mid F_K(q^2)\mid^2$ & $q^2$ & $\mid F_K(q^2)\mid^2$ \\
${\rm (GeV/c)}^2$ & & ${\rm (GeV/c)}^2$ & \\ \hline
0.0244 & $1.071\pm 0.083$ & 0.0744 & $1.242\pm 0.082$ \\
0.0294 & $1.086\pm 0.083$ & 0.0794 & $1.257\pm 0.082$ \\
0.0344 & $1.102\pm 0.083$ & 0.0844 & $1.280\pm 0.081$ \\
0.0394 & $1.118\pm 0.083$ & 0.0894 & $1.300\pm 0.081$ \\
0.0444 & $1.134\pm 0.083$ & 0.0944 & $1.316\pm 0.081$ \\
0.0494 & $1.150\pm 0.082$ & 0.0994 & $1.344\pm 0.081$ \\
0.0544 & $1.169\pm 0.082$ & 0.1044 & $1.364\pm 0.081$ \\
0.0594 & $1.187\pm 0.082$ & 0.1094 & $1.396\pm 0.081$ \\
0.0644 & $1.204\pm 0.082$ & 0.1144 & $1.413\pm 0.081$ \\
0.0694 & $1.222\pm 0.082$ & 0.1194 & $1.431\pm 0.082$ \\ \hline
\end{tabular}
}
\end{table}
\end{center}
 
\vspace{0.5cm}
 
\noindent {\bf 3. The Linear $\sigma$ Model}
 
\vspace{0.25cm}
 
The quark-level Linear $\sigma$ Model (L$\sigma$M) has been
discussed in several
papers (e.g. \cite{gml,MDS_FF,mike2,delmds,pavmds}). 
Scadron {\it et al.}\cite{MDS_FF} gives the L$\sigma$M
expressions for meson form factors in the chiral limit (CL),
 
\begin{eqnarray}
F(q^2)^{CL}_{L\sigma M}=-4ig^2N_c/(16\pi^4)\int_0^1dx\int d^4p~
[p^2-m^2+x(1-x)q^2]^{-2}.
\label{eq.7}
\end{eqnarray}
   
\noindent Here, $N_c$, the number of colours, is related to the 
meson-quark coupling, $g$, by \cite{MDS_FF,delmds}
 
\begin{eqnarray}
g=2\pi/\sqrt{N_c}.
\label{eq.8}
\end{eqnarray}
  
\noindent Thus equation (\ref{eq.7}) becomes
 
\begin{eqnarray}
F(q^2)^{CL}_{L\sigma M}=\frac{-i}{\pi^2}\int_0^1dx\int d^4p~
[p^2-m^2+x(1-x)q^2]^{-2}.
\label{eq.9}
\end{eqnarray}
  
\noindent In equations (\ref{eq.7}) and (\ref{eq.9}), setting $m$
equal to the mean up-down 
constituent quark mass, $\hat{m}=(m_u+m_d)/2$ gives the pion form
factor, $F_{\pi}$. For the kaon form factor, $m$ is set to the
mean of $\hat{m}$ and the s quark mass, $m_{us}=(m_s+\hat{m})/2$.
In the present paper we use values in the chiral limit 
\cite{comment}: 

\begin{eqnarray}
\hat{m}^{CL}=325.7~{\rm MeV},~~~ m_s^{CL}=469.0~{\rm MeV}, ~~~ 
m_{us}^{CL}=397.4~{\rm MeV}.
\label{eq.10}
\end{eqnarray}
 
To avoid the divergence in equation (\ref{eq.9}), we use a
Taylor-series expansion
as follows. Differentiating equation (\ref{eq.9}) gives
 
\begin{eqnarray}
\frac{dF(q^2)^{CL}_{L\sigma M}}{dq^2}=
\frac{2i}{\pi^2}\int_0^1dx\int \frac{d^4p}
{[p^2-m^2+x(1-x)q^2]^3}.
\label{eq.11}
\end{eqnarray}

\noindent The second integral in equation (\ref{eq.11}) can be
written

\begin{eqnarray}
I\equiv\int \frac{d^4p}{[p^2-V]^3}
\label{eq.12}
\end{eqnarray}

\noindent where $V=m^2-x(1-x)q^2$. Applying the Wick rotation,
$k^2=-p^2$, $d^4p=i\pi^2k^2dk^2$, this becomes

\begin{eqnarray}
I = \int \frac{i\pi^2k^2~dk^2}{[-k^2-V]^3}
\label{eq.13}
\end{eqnarray} 

\noindent which can be integrated directly to give

\begin{eqnarray}
I =-\frac{i\pi^2}{2V}
\label{eq.14}
\end{eqnarray}

\noindent and hence, from equation (\ref{eq.11}),

\begin{eqnarray}
\frac{dF(q^2)^{CL}_{L\sigma M}}{dq^2}=
\int_0^1 \frac{x(1-x)}{V}dx.
\label{eq.15}
\end{eqnarray}

\noindent As a check, the integral in equation (\ref{eq.12})
can be evaluated
from the expression given by Scadron \cite{aqt}, and also by 
integration of the Taylor series for the charge radius given
in \cite{bels}.
 
At $q^2=0$, $V=m^2$ gives

\begin{eqnarray}
\bigg[\frac{dF(q^2)^{CL}_{L\sigma M}}{dq^2}\bigg]_{q^2=0}=
\frac{1}{m^2}\int_0^1 x(1-x)dx=\frac{1}{6m^2}.
\label{eq.16}
\end{eqnarray}
  
\noindent Further derivatives follow from differentiating 
equation (\ref{eq.15}).
In writing a Taylor series for $F(q^2)$, the first term, 
$F(0)^{CL}_{L\sigma M}$, is given directly by the normalisation
requirement that $F(q^2)=1$ at $q^2=0$. The series, therefore, is
 
\begin{eqnarray}
F(q^2)^{CL}_{L\sigma M}
=1+\frac{1}{6}\bigg(\frac{q}{m}\bigg)^2+
\frac{1}{60}\bigg(\frac{q}{m}\bigg)^4+
\frac{1}{420}\bigg(\frac{q}{m}\bigg)^6+
\frac{1}{2520}\bigg(\frac{q}{m}\bigg)^8+
\frac{1}{13860}\bigg(\frac{q}{m}\bigg)^{10}+...
\label{eq.17}
\end{eqnarray}
 
\noindent As in equation (\ref{eq.9}), this series gives 
$F_{\pi}(q^2)$ with $m=\hat{m}$ and $F_K(q^2)$ with $m=m_{us}$.
 
The form factors $F_{\pi}(q^2)^{CL}_{L\sigma M}$ and 
$F_K(q^2)^{CL}_{L\sigma M}$ predicted by equation (\ref{eq.17})
are plotted in figures \ref{fig.2} and \ref{fig.3}, together with
the data from the present analysis and from \cite{radii}.
For both $F_{\pi}(q^2)^{CL}_{L\sigma M}$ and 
$F_K(q^2)^{CL}_{L\sigma M}$, the L$\sigma$M shows the same
qualitative trend as the data, but falls somewhat below the
experimental points. However, equation (\ref{eq.7}), from
\cite{MDS_FF}, is derived using quark loops only. Several papers
(e.g. \cite{tarr,brs}) have studied the addition of meson loops
to these. Although they do not affect the result at $q^2=0$, they
may be important away from this point, which may be the source
of the discrepancy. 

\vspace{0.5cm}
 
\noindent{\bf 4. Quark triangle graphs}
  
\vspace{0.25cm}

\begin{figure}[t]
\hspace{1.5in}
\epsfig{file=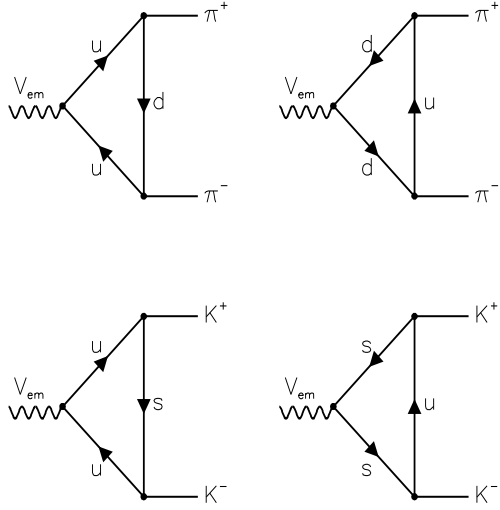,height=10.0cm,width=9.0cm}
\caption{\label{fig.4} Quark-triangle graphs.}
\end{figure}

Scadron {\it et al} \cite{MDS_FF} describe another approach to 
meson form factors, {\it via} the quark-triangle graphs of figure
\ref{fig.4}. From \cite{MDS_FF}, the pion and kaon form factors are
given by

\begin{eqnarray}
F_{\pi}(q^2)_{QT}=-4ig^2N_c\bigg[
\frac{2}{3}I(q^2,m_u^2,m_d^2,m_{\pi}^2)+
\frac{1}{3}I(q^2,m_d^2,m_u^2,m_{\pi}^2)\bigg],
\label{eq.18}
\end{eqnarray}

\begin{eqnarray}
F_K(q^2)_{QT}=-4ig^2N_c\bigg[
\frac{2}{3}I(q^2,m_u^2,m_s^2,m_K^2)+
\frac{1}{3}I(q^2,m_s^2,m_u^2,m_K^2)\bigg],
\label{eq.19}
\end{eqnarray}
 
\noindent where
 
\begin{eqnarray}
I(q^2,m_1^2,m_2^2,M^2)&=&
\frac{i\pi^2}{2(2\pi)^4}
\int_0^1dv\int_v^1du\frac{q^2u+2(M^2-(m_1-m_2)^2)(1-u)}
{m_2^2-(M^2+m_2^2-m_1^2)u+M^2u^2+(v^2-u^2)q^2/4}
\nonumber \\
 &+& \int_0^1dx\int d^4p
\bigg[p^2-m_2^2+(M^2+m_2^2-m_1^2)x-M^2x^2\bigg]^{-2}.
\label{eq.20}
\end{eqnarray}
 
\noindent Here, we replace $m_u$ and $m_d$ by $\hat{m}$ from
equation (\ref{eq.10}). These expressions can be calculated
either by direct evaluation of the integral in equation 
(\ref{eq.20}) or, as a check, by construction of a
Taylor series by repeated differentiation of equation (\ref{eq.20}). 

The results for the pion and kaon form factors are shown in
figures \ref{fig.2} and \ref{fig.3}. As for the L$\sigma$M, the
predictions for both $F_K$ and $F_{\pi}$ follow the data
qualitatively, but detailed quantitative agreement is lacking.
As with the L$\sigma$M, the problem may lie with the neglect of
meson loops, which could be significant when $q^2>0$.

\vspace{0.5cm}

\noindent{\bf 5. Vector meson dominance}
  
\vspace{0.25cm}

Finally, we examine a simple model based on VMD. As discussed in
section 2 above, the pion form factor is dominated in this model by
the $\rho$ pole, equation (\ref{eq.5}), and figure \ref{fig.2}
shows that this simple picture agrees well with the data.

For the kaon form factor, there are contributions from
the $\rho$, $\omega$ and $\phi$ poles:
 
\begin{eqnarray}
F_K(q^2)_{VMD}=N\bigg(\frac{1}{2}
\frac{g_{\rho ee}}{m^2_{\rho^0}-q^2}
+\frac{1}{2}\frac{g_{\omega ee}}{m^2_{\omega}-q^2}+
\sqrt{\frac{1}{2}}\frac{g_{\phi ee}}{m^2_{\phi}-q^2}\bigg).
\label{eq.21}
\end{eqnarray}
 
\noindent where $g_{\rho ee}=4.97$, $g_{\omega ee}=17.06$ and 
$g_{\phi ee}=13.38$, derived from the decay widths. The
$\rho^0K^+K^-$, $\omega K^+K^-$ and $\phi K^+K^-$ SU(3) 
coefficients are $1/2$, $1/2$ and 
$1/sqrt{2}$ respectively. The requirement that $F(0)=1$
gives the normalisation coefficient as $N=0.03682~{\rm GeV}^2$. 
 
The prediction of equation (\ref{eq.21}) is plotted with the data
for $F_K$ in figure \ref{fig.3}. As for $F_{\pi}$, the VMD gives
excellent agreement with
data, both for the previously available data for $q^2<0$ and
for the new data derived in the present paper.
 
\vspace{0.5cm}
 
\noindent{\bf 6. Slopes of form factors at $q^2=0$.}

\vspace{0.25cm}

In this section we examine two properties that depend on the
slopes of the form factors at $q^2=0$, i.e. the meson charge radii
and the amplitude for $K^+\rightarrow\pi^+e^+e^-$ at $q^2=0$.
Predictions for charge radii have been published before \cite{MDS_FF}
but we include them here for completeness and because we update
some numerical values.
 
The charge radius is related to the form factor by \cite{MDS_FF}
 
\begin{eqnarray}
r\equiv\sqrt{\langle r^2\rangle}=
\sqrt{6\bigg[\frac{dF(q^2)}{dq^2}\bigg]_{q^2=0}}.
\label{eq.22}
\end{eqnarray}

\noindent For the L$\sigma$M and VMD, the quantity $dF(q^2)/dq^2$
is straightforwardly obtained from  equations (\ref{eq.16}),
(\ref{eq.5}) and (\ref{eq.21}). For the QT model, the expression
is given by \cite{MDS_FF} as
 
\begin{eqnarray}
\langle r_{\pi}^2\rangle=\frac{g^2N_c}{4\pi^2\hat{m}^2}
=\bigg[\frac{\hbar c}{\hat{m}}\bigg]^2
\label{eq.23}
\end{eqnarray}

\noindent and

\begin{eqnarray}
\langle r_K^2\rangle=
\frac{g^2N_c}{4\pi^2\hat{m}^2}\bigg(1-\frac{5}{6}\delta+
\frac{3}{5}\delta^2-\frac{4}{9}\delta^3+
\frac{22}{63}\delta^4-\frac{2}{7}\delta^5+...\bigg).
\label{eq.24}
\end{eqnarray}

\noindent where $\delta=(m_s/\hat{m})-1=0.44$ (see also 
\cite{aybram}).

The predictions are compared with the experimental values from
\cite{PDG} in table \ref{tab.2}.

\begin{center}
\begin{table}
\caption{\label{tab.2} Meson charge radii and amplitudes for
$K^+\rightarrow\pi^+e^+e^-$ at $q^2=0$.}
\vspace{0.3cm}
\center{
\begin{tabular}{|l|c|c|c|}   
\hline
Model     & $r_{\pi}$      & $r_K$   &     $A(0)$         \\
          &   (fm)      &  (fm)   &$10^{-9}~{\rm GeV}^{-2}$\\ \hline
L$\sigma$M&  0.606         & 0.497       &         7.56   \\
QT        &  0.606         & 0.514       &         6.47   \\
VMD       &  0.623         & 0.574       &         3.69   \\
Experiment&$0.672\pm 0.008$&$0.56\pm 0.03$&$4.00 \pm 0.18$\\ \hline
\end{tabular}
}
\end{table}
\end{center}

The LD contribution to the amplitude for
$K^+\rightarrow\pi^+e^+e^-$ at $q^2=0$ is given by equation
(\ref{eq.1}) as

\begin{eqnarray}
\mid A_{LD}(0)\mid=e^2\bigg|
\frac{\langle\pi^+\mid H_W\mid 
K^+\rangle}{m^2_{K^+}-m^2_{\pi^+}}\bigg|~~\bigg|
\frac{dF_{\pi^+}}{dq^2}-\frac{dF_{K^+}}{dq^2}\bigg|_{q^2=0}.
\label{eq.25}
\end{eqnarray}
 
\noindent Since the pion loop term vanishes at $q^2=0$, equation
(\ref{eq.25}) gives the total amplitude directly. For the three
models considered here, $A(0)$ can be calculated using the
expressions for the derivatives given above. We take the 
experimental value from Brookhaven E865 \cite{865}, converting
their $f_0$ to our $A(0)$ using equation (\ref{eq.4}).

From table \ref{tab.2}, we see that the VMD picture again comes
closest
to a quantitative description of the data. As expected, this is
consistent with the conclusion from the form factors of figures
\ref{fig.2} and \ref{fig.3}.

\vspace{0.5cm}
 
\noindent{\bf 7. Summary}
 
\vspace{0.25cm}
 
We have derived new experimental values for $F_K$ for positive
$q^2$ from the decay $K^+\rightarrow\pi^+e^+e^-$, extending the
upper limit of the range over which experimental values of
$F_K(q^2)$ are known to $q^2=0.125~({\rm GeV/c})^2$. Also, we
have calculated both $F_{\pi}$ and $F_K$ for the range of $q^2$
covered by these new data and the previous data, and have examined
the slopes at $q^2=0$. It is apparent that for all these
comparisons, the L$\sigma$M and the QT model give at best a
qualitative description of the data, probably due to the neglect
of meson loops. By contrast, the VMD model gives a detailed
quantitative fit both to the data for $F_{\pi}$ and to the new
data for $F_K$.
 
\vspace{0.5cm}
               
\noindent{\bf Acknowledgments}
 
\vspace{0.25cm}
 
We are grateful to F. Kleefeld for useful comments. We acknowledge
support from the US DOE.


\begin{thebibliography}{99}
\bibitem{comment}
We derive values for the quark masses in the chiral limit as follows.
The Goldberger-Treiman relation \cite{MDS_FF,nsh} 
$\hat{m}^{CL}=f_{\pi}^{CL}g=325.7~{\rm MeV}$ for $g=2\pi/\sqrt{3}$
from reference 14 of \cite{nsh} due to the once-subtracted
dispersion relation for $f_{\pi}$ (see equation (19) of
\cite{nsh}). Also \cite{PDG,MDS_FF,nsh}, 
$m_s/\hat{m}=2f_K/f_{\pi}-1=1.44$ so that $m_s^{CL}=469.0~{\rm MeV}$.
\end{thebibliography}
\end{document}